\documentclass[preprint2]{aastex}
\shorttitle{NGC 2024 FIR 6 Water and Methanol Masers}
\shortauthors{Choi, Lee, Byun, and Kang}

\usepackage{times}
\slugcomment{To appear in the Astrophysical Journal}

\begin{document}

\fontsize{10}{10.6}\selectfont

\title{Water and Methanol Maser Activities in the NGC 2024 FIR 6 Region}
\author{\sc Minho Choi$^{1}$, Miju Kang$^1$, Do-Young Byun$^1$,
        and Jeong-Eun Lee$^2$}
\affil{$^1$ Korea Astronomy and Space Science Institute,
            776 Daedeokdaero, Yuseong, Daejeon 305-348, Republic of Korea;
            minho@kasi.re.kr \\
       $^2$ Department of Astronomy and Space Science,
            Kyung Hee University, Yongin, Gyeonggi 446-701, Republic of Korea}
\setcounter{footnote}{2}

\begin{abstract}
\fontsize{10}{10.6}\selectfont
The NGC 2024 FIR 6 region was observed in the water maser line at 22 GHz
and the methanol class I maser lines at 44, 95, and 133 GHz.
The water maser spectra displayed
several velocity components and month-scale time variabilities.
Most of the velocity components may be associated with FIR 6n,
while one component was associated with FIR 4.
A typical life time of the water-maser velocity-components is about 8 months.
The components showed velocity fluctuations
with a typical drift rate of about 0.01 km s$^{-1}$ day$^{-1}$.
The methanol class I masers were detected toward FIR 6.
The methanol emission is confined
within a narrow range around the systemic velocity of the FIR 6 cloud core.
The methanol masers suggest the existence of shocks
driven by either the expanding H {\small II} region of FIR 6c
or the outflow of FIR 6n.
\end{abstract}


\section{INTRODUCTION}

The NGC 2024 region is a site of star formation
at a distance of 415 pc from the Sun (Anthony-Twarog 1982).
FIR 6 is one of the dense cores in the NGC 2024 molecular ridge (FIR 1--7),
first identified by observations in the dust continuum (Mezger et al. 1988),
and was known to be associated with an H$_2$O maser source
and a molecular outflow (Genzel \& Downes 1977; Richer 1990).
Studies with the millimeter continuum and molecular lines suggested
that FIR 6 may contain a protostar
(Schulz et al. 1991; Chandler \& Carlstrom 1996; Wiesemeyer et al. 1997;
Visser et al. 1998; Mangum et al. 1999).

High-resolution imaging observations revealed a binary system
(Lai et al. 2002).
The projected separation between the binary members, FIR 6c and 6n,
is 3\farcs9 or 1600 AU,
and both of them are associated with H$_2$O maser sources (Choi et al. 2012).
The primary (in the millimeter continuum), FIR 6c,
has a flat continuum spectrum around 1 mm
and suggested to be a hypercompact H {\small II} region
probably powered by a B1 star (Choi et al. 2012).
The H$_2$O maser line of FIR 6c detected in 1996 was blueshifted
by $\sim$20 km s$^{-1}$ with respect to the ambient cloud velocity,
suggesting the existence of an outflow activity (Choi et al. 2012).
The secondary, FIR 6n, exhibits relatively stronger H$_2$O maser emission
and drives a bipolar outflow
(Furuya et al. 2003; Alves et al. 2011; Choi et al. 2012).
FIR 6n may be a low-mass young stellar object (YSO),
probably an active protostar (Choi et al. 2012).
As the observed phenomena displayed by FIR 6
are a mixture of high- and low-mass star formation activities,
further investigations are needed to understand this interesting region.

In this paper, we present the results of
our observations of the NGC 2024 FIR 6 region
in the H$_2$O, CH$_3$OH, and $^{13}$CS lines
with the Korean Very Long Baseline Interferometry Network (KVN) antennas.
We describe our spectral line observations in Section 2.
In Section 3, we report the results.
In Section 4, we discuss the star-forming activities
in the FIR 6 and 4 regions.
A summary is given in Section 5.

\section{OBSERVATIONS}

The NGC 2024 FIR 6 region was observed using the KVN 21 m antennas
in the single-dish telescope mode
during the 2011--2012 observing season.
The observations were carried out
with the KVN Yonsei telescope at Seoul, Korea
in 2011 November and 2012 January,
and the KVN Ulsan telescope at Ulsan, Korea in 2012 April-May.

The data were obtained
using the high-electron-mobility-transistor receivers
in the 22, 44, and 86 GHz bands
and the superconductor-insulator-superconductor receivers
in the 129 GHz band.
Digital filters and spectrometers were used as the back end (Lee et al. 2011).
The spectrometers were used
in the mode of 4 intermediate-frequency data streams.
Each intermediate-frequency stream was configured
to have a band width of 64 MHz and 4096 channels,
which gives a spectral channel width of 15.625 kHz.
The antenna temperature was calibrated by the standard chopper-wheel method,
which automatically corrected for the effects of atmospheric attenuation.

\begin{deluxetable}{lcc}
\tabletypesize{\small}
\tablecaption{\small Parameters of Detected Maser Lines}%
\tablewidth{0pt}
\tablehead{
& \colhead{Frequency\tablenotemark{a}} & \\
Line & \colhead{(GHz)} & \colhead{Observing Run}}%
\startdata
H$_2$O $6_{16} \rightarrow 5_{23}$       & \phn22.2350771
         & 2011 Nov 6, 2012 Jan 27, 2012 Apr 12--14, 2012 May 11--12 \\
CH$_3$OH $7_{0} \rightarrow 6_{1}$ $A^+$ & \phn44.069476\phn
         & 2012 Apr 12--15, 2012 May 11--13 \\
CH$_3$OH $8_{0} \rightarrow 7_{1}$ $A^+$ & \phn95.169516\phn
         & 2012 Jan 27, 2012 Apr 12--15 \\
CH$_3$OH $6_{-1} \rightarrow 5_{0}$ $E$  &    132.890800\phn
         & 2012 Apr 12--15 \\
\enddata\\
\tablenotetext{a}{Rest frequency from the line catalogue
                  of the National Institute of Standards and Technology (NIST)
                  recommended rest frequencies
                  for observed interstellar molecular microwave transitions
                  by F. J. Lovas
                  (http://www.nist.gov/pml/data/micro/index.cfm).}%
\end{deluxetable}

\begin{deluxetable}{lcccccc}
\tabletypesize{\small}
\tablecaption{\small Telescope Parameters}
\tablewidth{0pt}
\tablehead{
&& \colhead{Frequency\tablenotemark{a}} & \colhead{Beam\tablenotemark{b}}
&&& \colhead{$f$\tablenotemark{d}} \\
Telescope & \colhead{Observing Run} & \colhead{(GHz)}
& \colhead{(arcsec)} & \colhead{$\eta_A$\tablenotemark{c}}
& \colhead{$\eta_{\rm mb}$\tablenotemark{c}} & \colhead{(Jy K$^{-1}$)}}%
\startdata
KVN Yonsei & 2011 Nov, 2012 Jan & \phn22 &    118 & 0.65 & 0.46 & 15.3 \\
           &                    & \phn86 & \phn28 & 0.48 & 0.36 & 20.7 \\
KVN Ulsan  & 2012 Apr, 2012 May & \phn22 &    119 & 0.62 & 0.44 & 16.0 \\
           &                    & \phn44 & \phn62 & 0.62 & 0.47 & 16.0 \\
           &                    & \phn86 & \phn29 & 0.49 & 0.39 & 20.3 \\
           &                    &    129 & \phn22 & 0.32 & 0.29 & 38.8 \\
\enddata\\
\tablenotetext{a}{Frequency where the beam size and efficiencies
                  were measured.}%
\tablenotetext{b}{Full width at half maximum (FWHM) of the main beam.}%
\tablenotetext{c}{Aperture and main-beam efficiencies.
                  In each receiver band, $\eta_A$ is nearly constant,
                  and $\eta_{\rm mb}$ is inversely proportional
                  to the square of frequency.}%
\tablenotetext{d}{Scaling factor for converting the KVN raw data
                  to spectra in the flux density scale
                  for the maser lines in Table 1,
                  which includes the telescope sensitivity,
                  quantization correction factor (1.25),
                  and sideband separation efficiency
                  (0.8 for the KVN Ulsan 129 GHz band and 1.0 otherwise).}%
\end{deluxetable}

The main target source was NGC 2024 FIR 6
at $\alpha_{2000}$ = 05$^{\rm h}$41$^{\rm m}$45\fs14,
$\delta_{2000}$ = --01\arcdeg56$'$04\farcs4,
which corresponds to the 6.9 mm continuum source FIR 6c (Choi et al. 2012).
The data were taken by position switching
with the off position of
(05$^{\rm h}$43$^{\rm m}$05\fs19, --01\arcdeg56$'$04\farcs4),
which is a nearby empty area
in the CS $J$ = 2 $\rightarrow$ 1 map of Lada et al. (1991).

Telescope pointing was checked by observing Orion IRc2 (Baudry et al. 1995)
in the SiO $v$ = 1 $J$ = 1 $\rightarrow$ 0 maser line.
The pointing observations were performed about once in an hour.
The telescope pointing was good within $\sim$4$''$.

\subsection{Main Target Lines}

The main target lines were
the H$_2$O $6_{16} \rightarrow 5_{23}$ (22 GHz) line
and the CH$_3$OH $7_{0} \rightarrow 6_{1}$ $A^+$,
$8_{0} \rightarrow 7_{1}$ $A^+$, and $6_{-1} \rightarrow 5_{0}$ $E$
(44, 95, and 133 GHz, respectively) lines.
These CH$_3$OH lines are class I maser lines.
Table 1 lists the parameters of these lines.
The spectrometer setting gives a velocity channel width of
0.21 km s$^{-1}$ for the H$_2$O line
and 0.11 km s$^{-1}$ for the CH$_3$OH 44 GHz line,
and we present the data of these lines without further resampling.
For the CH$_3$OH 95 GHz line,
the spectra were smoothed using a Hanning window over three channels,
which gives a velocity channel width of 0.098 km s$^{-1}$.
For the CH$_3$OH 133 GHz line,
the Hanning smoothing was applied twice,
which gives a velocity channel width of 0.14 km s$^{-1}$.

In the 2012 May run,
a map was made in the H$_2$O line,
covering a 4\farcm8 $\times$ 4\farcm8 region around the FIR 1--7 cores.
The mapping grid size was
12$''$ near FIR 3--7 and 48$''$ in the periphery of the map.
Some extra integrations were made toward FIR 4
at (05$^{\rm h}$41$^{\rm m}$44\fs14, --01\arcdeg54$'$46\farcs0).

The data were calibrated
using the standard efficiencies of KVN for the 2011--2012 observing season
(Table 2; also see Lee et al. 2011 and http://kvn-web.kasi.re.kr/).
We will present the maser data in the flux density scale.
For each spectrum, a first-order baseline was removed.
The baseline was determined from the velocity intervals
of $V_{\rm LSR}$ = (--10, 0) and (20, 30) km s$^{-1}$ for the H$_2$O line
and from (1, 6) and (16, 21) km s$^{-1}$ for the CH$_3$OH lines.

\subsection{Other Methanol Lines}

In addition to the class I maser lines described above,
FIR 6 was observed in most of the known CH$_3$OH lines
in the tuning ranges of the KVN receivers.
These observations were carried out in the 2012 April--May observing runs.
Table 3 lists the line parameters.
None of these lines was detected.

\begin{deluxetable}{lccc}
\tabletypesize{\small}
\tablecaption{\small Parameters of Undetected Methanol Lines}
\tablewidth{0pt}
\tablehead{
& \colhead{Frequency\tablenotemark{a}} & \colhead{rms\tablenotemark{b}}
& \colhead{rms\tablenotemark{c}} \\
Line & \colhead{(GHz)}         & \colhead{(K)} & \colhead{(Jy)}}%
\startdata
CH$_3$OH $12_{2}  \rightarrow 11_{1}$ $A^-$ $v_t = 1$
                                            & \phn21.550342    & 0.16 &\ldots \\
CH$_3$OH $9_{2}   \rightarrow 10_{1}$ $A^+$ & \phn23.121024    & 0.14 & 0.7   \\
CH$_3$OH $6_{-2}  \rightarrow 7_{-1}$ $E$   & \phn85.568074    & 0.27 & 1.7   \\
CH$_3$OH $7_{2}   \rightarrow 6_{3}$ $A^-$  & \phn86.615602    & 0.24 & 1.5   \\
CH$_3$OH $7_{2}   \rightarrow 6_{3}$ $A^+$  & \phn86.902947    & 0.21 & 1.3   \\
CH$_3$OH $15_{3}  \rightarrow 14_{4}$ $A^+$ & \phn88.594809    & 0.23 &\ldots \\
CH$_3$OH $15_{3}  \rightarrow 14_{4}$ $A^-$ & \phn88.939993    & 0.21 &\ldots \\
CH$_3$OH $8_{-4}  \rightarrow 9_{-3}$ $E$   & \phn89.505778    & 0.31 &\ldots \\
CH$_3$OH $20_{-3} \rightarrow 19_{-2}$ $E$ $v_t = 1$
                                            & \phn90.81239\phn & 0.29 &\ldots \\
CH$_3$OH $8_{3}   \rightarrow 9_{2}$ $E$    & \phn94.541806    & 0.31 & 1.6   \\
CH$_3$OH $6_{0}   \rightarrow 5_{1}$ $E$    &    124.569976    & 0.45 &\ldots \\
CH$_3$OH $12_{1}  \rightarrow 11_{2}$ $A^-$ &    129.433406    & 0.41 &\ldots \\
CH$_3$OH $6_{2}   \rightarrow 7_{1}$ $A^-$  &    132.621859    & 0.43 & 3.0   \\
CH$_3$OH $5_{-2}  \rightarrow 6_{-1}$ $E$   &    133.605385    & 0.87 & 5.9   \\
CH$_3$OH $12_{-3} \rightarrow 13_{-2}$ $E$  &    134.231013    & 0.37 &\ldots \\
CH$_3$OH $7_{-4}  \rightarrow 8_{-3}$ $E$   &    137.902997    & 0.41 &\ldots \\
CH$_3$OH $23_{-2} \rightarrow 23_{1}$ $E$   &    140.03314\phn & 0.57 &\ldots \\
CH$_3$OH $18_{0}  \rightarrow 18_{-1}$ $E$  &    140.151188    & 0.38 &\ldots \\
CH$_3$OH $0_{0}   \rightarrow 1_{1}$ $E$ $v_t = 1$
                                            &    141.441249    & 0.52 &\ldots \\
\enddata\\
\tablenotetext{a}{Rest frequency from the NIST catalogue.}%
\tablenotetext{b}{Noise rms level in the $T_{\rm mb}$ scale,
                  measured from spectra with a channel width
                  of 15.625 kHz (0.2 km s$^{-1}$) for the 22 GHz band,
                  31.25 kHz ($\sim$0.11 km s$^{-1}$) for the 86 GHz band,
                  and 62.5 kHz ($\sim$0.14 km s$^{-1}$) for the 129 GHz band.}%
\tablenotetext{c}{Noise rms level in the flux density scale
                  for class II maser lines
                  (Menten 1991; Cragg et al. 1992; M{\"u}ller et al. 2004).}%
\end{deluxetable}

\subsection{The $^{13}$CS $J$ = 2 $\rightarrow$ 1 Line}

FIR 6 was also observed
in the $^{13}$CS $J$ = 2 $\rightarrow$ 1 line (92.494270 GHz)
in the 2012 April observing run.
The spectra were smoothed using a Hanning window over three channels,
which gives a velocity channel width of 0.10 km s$^{-1}$.
For each spectrum, a first-order baseline was removed.
The spectral baseline was determined from the velocity intervals
of (1, 6) and (16, 21) km s$^{-1}$.
We will present the $^{13}$CS data
in the main-beam temperature ($T_{\rm mb}$) scale.

\section{RESULTS}

\subsection{The Water Maser}

The H$_2$O maser line was detected in all the four observing runs.
The spectra showed several velocity components,
and each of them displayed significant month-scale time-variability.
Figure 1 shows the spectra from each observing run.

\begin{figure}[!t]
\epsscale{1.0}
\plotone{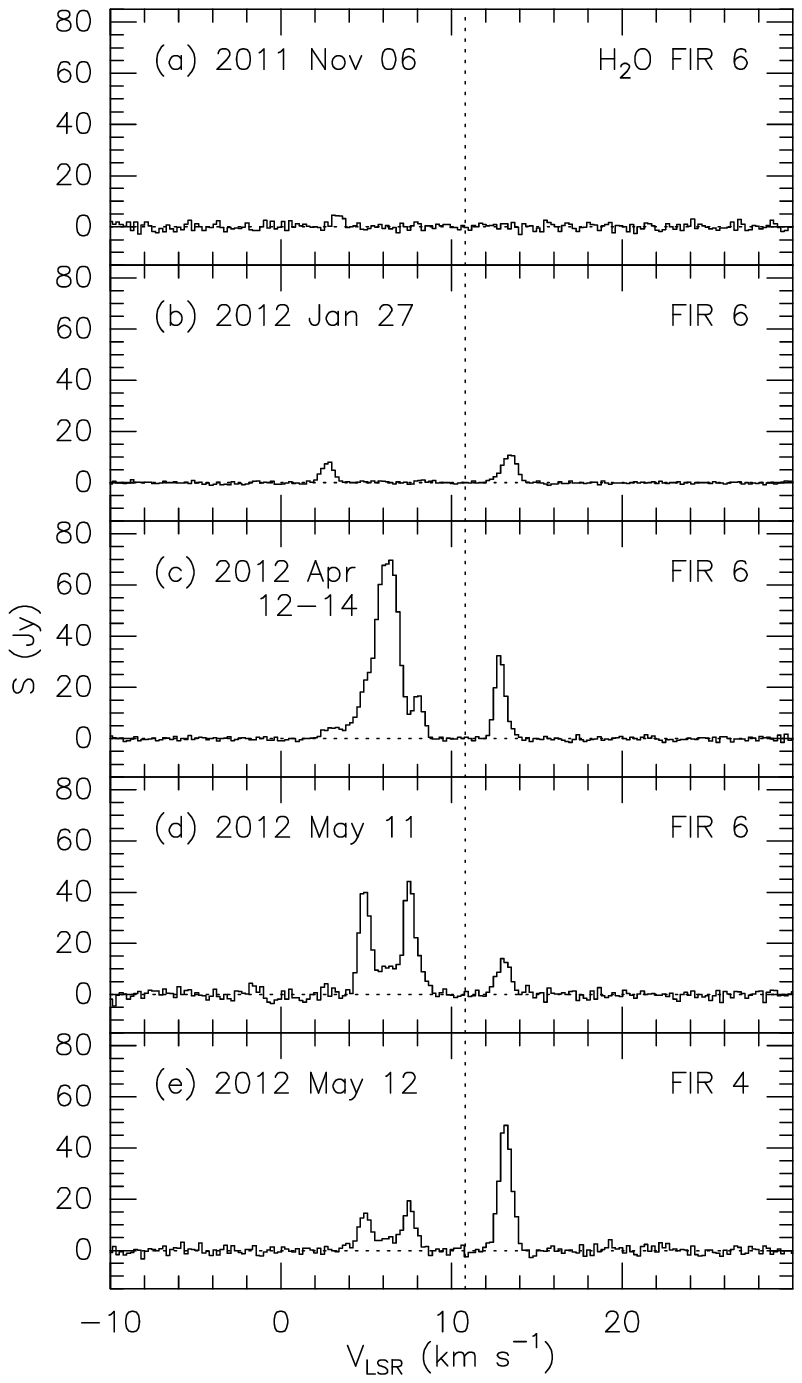}
\caption{\small\baselineskip=0.825\baselineskip
Spectra of the H$_2$O 22 GHz maser line.
(a--d)
Spectra toward FIR 6.
(e)
Spectrum toward FIR 4.
The observation dates are labeled.
Vertical dotted line:
velocity of the ambient dense gas
($V_{\rm LSR}$ = 10.8 km s$^{-1}$; Section 3.3).}
\end{figure}

Inspecting the spectra closely,
five velocity components can be identified:
$V_{\rm LSR} \approx$ 3, 5, 6, 8, and 13 km s$^{-1}$.
The 3 km s$^{-1}$ component was present in the first observing run,
became stronger in 2012 January, weakened in April, and then disappeared.
The 5 km s$^{-1}$ component was present in 2012 April
(as a shoulder of the 6 km s$^{-1}$ component),
and could be seen clearly in May.
The 6 km s$^{-1}$ component ``flared'' in 2012 April, and weakened in May.
The 8 km s$^{-1}$ component appeared in 2012 April,
and became stronger in May.
The 13 km s$^{-1}$ component appeared in 2012 January,
became stronger in April, and weakened in May.
The map data showed
that most of the velocity components are associated with FIR 6
while the 13 km s$^{-1}$ component is associated with FIR 4 (Section 3.1.1).
Table 4 lists the line-profile parameters
(centroid velocity, peak intensity, line FWHM, and integrated intensity)
derived by Gaussian fits to each velocity component.

\begin{deluxetable}{lcrcc}
\tabletypesize{\small}
\tablecaption{\small Parameters of the Water Maser Spectra}
\tablewidth{0pt}
\tablehead{
& \colhead{$V_0$\tablenotemark{a}} & \colhead{$S_p$}
& \colhead{$\Delta V$\tablenotemark{a}} & \colhead{$\int S dV$} \\
Run & \colhead{(km s$^{-1}$)} & \colhead{(Jy)}
& \colhead{(km s$^{-1}$)} & \colhead{(Jy km s$^{-1}$)}}%
\startdata
2011 Nov     & \phn3.34 &  5.1 & 0.72 & \phn3.9 $\pm$ 0.7 \\
2012 Jan     & \phn2.74 &  8.0 & 0.84 & \phn7.2 $\pm$ 0.3 \\
             &    13.41 & 11.1 & 1.03 &    12.1 $\pm$ 0.3 \\
2012 Apr     & \phn2.98 &  4.1 & 1.17 & \phn5.1 $\pm$ 0.5 \\
             & \phn5.38 & 21.1 & 1.85 &    41.5 $\pm$ 5.4 \\
             & \phn6.38 & 63.1 & 1.32 &    88.8 $\pm$ 5.4 \\
             & \phn8.09 & 16.9 & 0.64 &    11.4 $\pm$ 0.3 \\
             &    12.86 & 32.4 & 0.71 &    24.4 $\pm$ 0.3 \\
2012 May\tablenotemark{b}
             & \phn4.91 & 42.1 & 0.77 &    34.7 $\pm$ 1.3 \\
             & \phn6.19 & 11.5 & 1.06 &    12.9 $\pm$ 1.6 \\
             & \phn7.57 & 43.0 & 0.88 &    40.4 $\pm$ 1.3 \\
             &    13.17 & 49.4 & 0.89 &    46.6 $\pm$ 0.9 \\
\enddata\\
\tablenotetext{a}{Uncertainties of the centroid velocity and line FWHM
                  are much smaller than the channel width (0.21 km s$^{-1}$).}%
\tablenotetext{b}{The parameters of the 5, 6, and 8 km s$^{-1}$ components
                  are from the spectrum toward FIR 6 (Figure 1(d)),
                  and those of the 13 km s$^{-1}$ component
                  are from the spectrum toward FIR 4 (Figure 1(e)).}%
\end{deluxetable}

\subsubsection{Source Positions}

Examinations of the map data showed
that the 5, 6, and 8 km s$^{-1}$ components come from sources near FIR 6,
and they (if different) cannot be distinguished with the KVN beam.
By contrast, the 13 km s$^{-1}$ component
clearly comes from a different source near FIR 4.
(The 3 km s$^{-1}$ component had already disappeared,
and its relation to these sources is not clear.)
Figure 2 shows the H$_2$O line maps.
The intensity distribution in each panel is consistent
with what is expected from a point-like source (convolved with the beam).
To find the source position, each map was fitted
with a Gaussian intensity profile having the same FWHM as the beam.

\begin{figure}[!t]
\epsscale{1.0}
\plotone{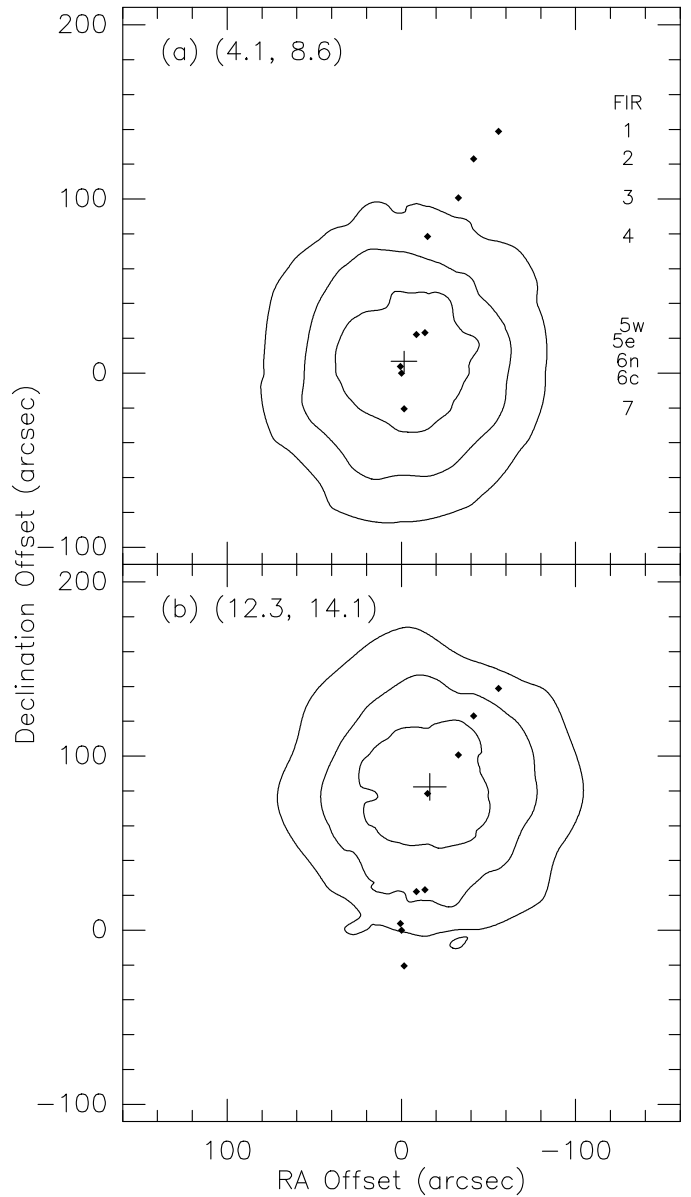}
\caption{\small\baselineskip=0.825\baselineskip
Maps of the H$_2$O 22 GHz maser emission in 2012 May.
The contour levels are 25, 50, and 75\%\ of the maximum value in each panel.
(a)
Map of the 5/6/8 km s$^{-1}$ components.
(b)
Map of the 13 km s$^{-1}$ component.
The velocity interval for the integration of intensity
is labeled in each panel.
Plus signs:
H$_2$O maser source positions derived by Gaussian fits.
The size of plus sign corresponds to the position uncertainty.
Diamond symbols:
millimeter continuum sources FIR 1--7
(Chandler \& Carlstrom 1996; Visser et al. 1998; Choi et al. 2012).}
\end{figure}

The best-fit source position of the 5/6/8 km s$^{-1}$ components
is (--2$''$ $\pm$ 7$''$, 7$''$ $\pm$ 6$''$) with respect to FIR 6c.
Among the two known H$_2$O maser sources in this region
(Choi et al. 2012; Furuya et al. 2003),
FIR 6n is located within the position uncertainty boundary,
and FIR 6c is just outside the boundary.%
\footnote{
As the description of the H$_2$O maser in Furuya et al. (2003)
is prone to confusion,
we reanalyzed the data set
of their observations with Very Large Array (project AF 354).
We found that the source coordinate in their Table 3 is correct,
and it corresponds to FIR 6n.
No H$_2$O maser emission was detected in the region around FIR 5.
The source described as
``7000 AU southeast of the close binary FIR 5'' in their Section 4.3.4
should be interpreted as FIR 6n.}
We presume that the source of the 5/6/8 km s$^{-1}$ components
is closely associated with FIR 6n,
while FIR 6c cannot be completely ruled out.

The best-fit source position of the 13 km s$^{-1}$ component
is (--1$''$ $\pm$ 9$''$, 4$''$ $\pm$ 8$''$) with respect to FIR 4.
There is no previous report of an H$_2$O maser detection in this region.
Since FIR 4 is the only known YSO within the uncertainty boundary,
we presume that the source of the 13 km s$^{-1}$ component
is associated with FIR 4.

Note that the first detection of the H$_2$O maser in the Orion B region
was reported by Johnston et al. (1973).
However, their position uncertainty was large (1$'$),
and FIR 3--6 are in the uncertainty boundary.
Therefore, it is difficult to identify the object
responsible for the H$_2$O emission detected by them.

\subsubsection{Variabilities}

Month-scale variabilities of the H$_2$O maser flux density
are shown in Figure 3(a).
Since four of the velocity components showed flux maxima
in the middle observing runs,
a rough estimate of timescale can be derived.
The FWHM of each light curve ranges from 60 to 180 days.
If the occurrence of each velocity component is a one-off event
(that is, shows a peak flux and disappears without a recurrence),
a typical life time (2FWHM) would be $\sim$8 months.

\begin{figure}[!b]
\epsscale{1.0}
\plotone{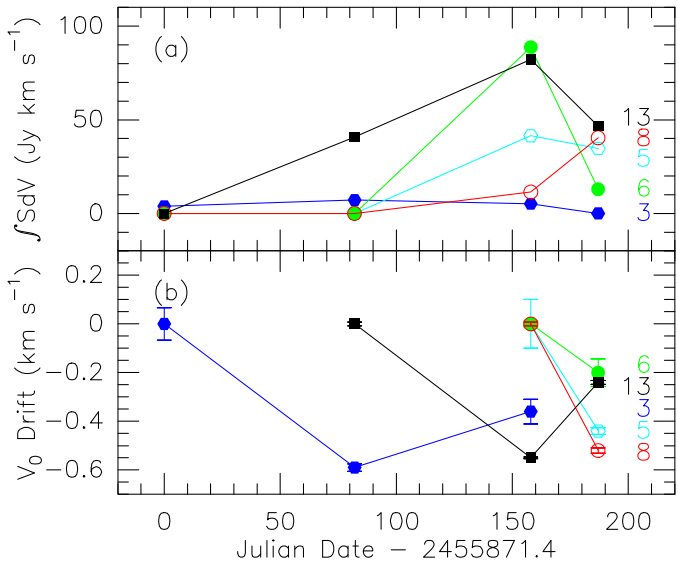}
\caption{\small\baselineskip=0.825\baselineskip
Variabilities of the H$_2$O 22 GHz maser for each velocity component.
The horizontal axis shows the time since the first observing run.
(a)
Light curve.
For the 3, 5, 6, and 8 km s$^{-1}$ components,
the integrated flux densities are from the spectra toward FIR 6.
For the 13 km s$^{-1}$ component,
the integrated flux densities of the 2012 January/April runs
are from the spectra toward FIR 6, corrected for the primary beam attenuation,
and that of the 2012 May run is from the spectrum toward FIR 4.
(b)
Time evolution of centroid velocity
with respect to the velocity at the first detection.
The velocities of the identified velocity components are labeled.}
\end{figure}

Figure 3(b) shows variabilities of the centroid velocity.
Interestingly, in all cases,
the velocity drifted to a lower value (that is, blueshifted)
in the first time interval after the appearance.
It is unlikely that this phenomenon is owing to an instrumental artifact
because the amount and sense of the drifts are all different
even in the same time interval.
(For example, in the last time interval of Figure 3(b),
the 13 km s$^{-1}$ component drifted to a higher velocity
while the others drifted to lower velocities at various rates.)
Those detected three times (the 3 and 13 km s$^{-1}$ components)
changed the sense of velocity drift.
Therefore, the centroid velocity is fluctuating
rather than drifting steadily.
A typical drift rate is $\sim$0.01 km s$^{-1}$ day$^{-1}$.
The velocity drift is usually interpreted
as the acceleration of emission source (e.g., Tofani et al. 1995),
but the fluctuations of the velocity seen in the FIR 4/6 masers
suggest that the cause of velocity variability is more complicated.

Day-scale variabilities were studied 
in the 2012 April observing run.
Figure 4 shows the change of spectra in two days.
A difference of line profile can be seen
in the 6 km s$^{-1}$ component only.
This component became stronger by 10\%\
and redshifted by $\sim$0.09 km s$^{-1}$.
(For comparison, the intensity variation of the 13 km s$^{-1}$ component
was smaller than 2\%.)
The information we can extract from the two spectra is rather limited.
Nevertheless, it is clear
that the sign of velocity drift in these two days
is the opposite of what was seen in a month
as describe in the paragraph above,
which reinforces the finding
that the velocity of emission source fluctuates.

\begin{figure}[!b]
\epsscale{1.0}
\plotone{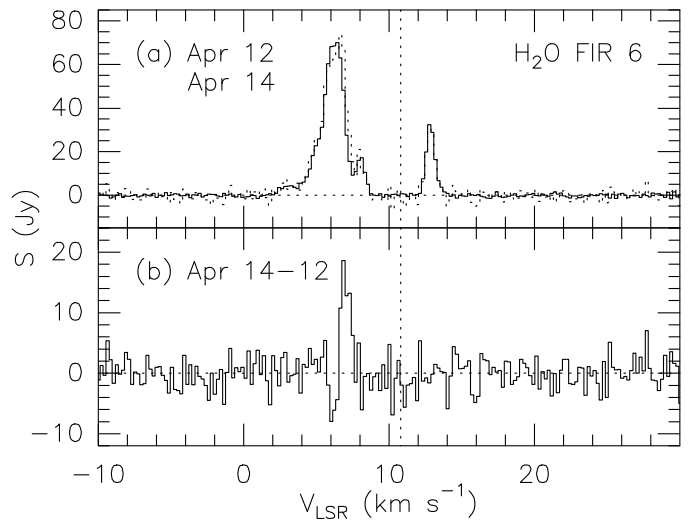}
\caption{\small\baselineskip=0.825\baselineskip
Daily spectra of the H$_2$O 22 GHz maser line toward FIR 6
in the 2012 April run.
(a)
Spectra on April 12 (solid line) and 14 (dotted line).
(b)
Difference between the two days.}
\end{figure}

\subsection{The Methanol Masers}

All the three CH$_3$OH class I maser lines
in the tuning ranges of the KVN receivers were detected toward FIR 6.
By contrast, other CH$_3$OH lines were undetected (Table 3),
which suggests that the detected lines are amplified emission.
Each of the 44 and 95 GHz lines was observed in two observing runs,
and there was no detectable time variability.

Figure 5 shows the spectra for each line,
and Table 5 lists the line-profile parameters.
All the detected CH$_3$OH emission is confined in a velocity interval
around the systemic velocity, within $\sim$2 km s$^{-1}$.
For the 44 GHz line, the residual spectra
of single-component and double-component Gaussian fits
show a peak at $\sim$10 km s$^{-1}$ higher than the noise level.
Therefore, the 44 GHz line seems to show three velocity components.
The signal-to-noise ratios of the 95 and 133 GHz lines
are too small to tell if there are multiple velocity components.

\begin{figure}[!t]
\epsscale{1.0}
\plotone{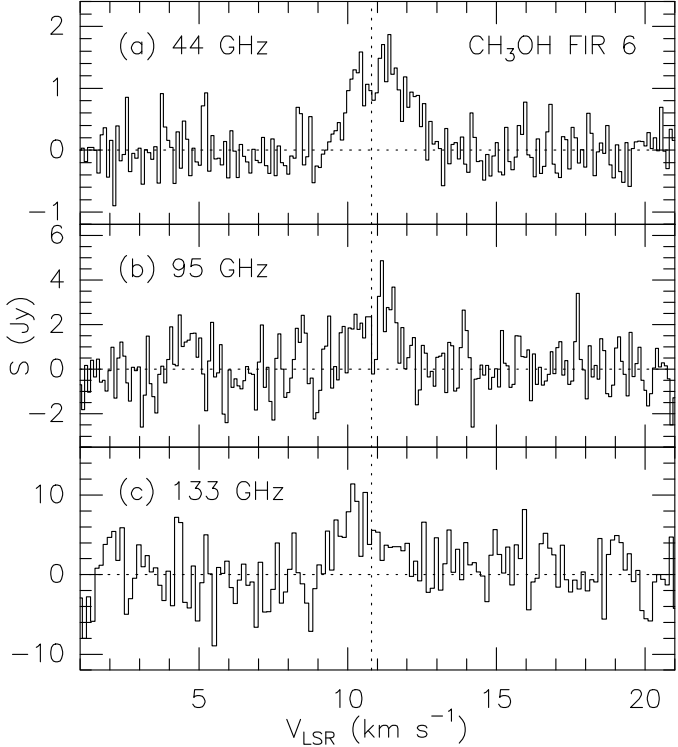}
\caption{\small\baselineskip=0.825\baselineskip
Spectra of the CH$_3$OH class I maser lines toward FIR 6.
(a)
CH$_3$OH $7_{0} \rightarrow 6_{1}$ $A^+$ line.
(b)
CH$_3$OH $8_{0} \rightarrow 7_{1}$ $A^+$ line.
(c)
CH$_3$OH $6_{-1} \rightarrow 5_{0}$ $E$ line.
For each line, all the data from the 2011--2012 season were averaged.}
\end{figure}

\begin{deluxetable}{lcccc}
\tabletypesize{\small}
\tablecaption{\small Parameters of the Methanol Maser Spectra}
\tablewidth{0pt}
\tablehead{
& \colhead{$V_0$} & \colhead{$S_p$} & \colhead{$\Delta V$}
& \colhead{$\int S dV$} \\
Line & \colhead{(km s$^{-1}$)} & \colhead{(Jy)} & \colhead{(km s$^{-1}$)}
& \colhead{(Jy km s$^{-1}$)}}%
\startdata
CH$_3$OH 44 GHz\tablenotemark{a}
               & 11.12 $\pm$ 0.08 & 1.4 & 2.2 $\pm$ 0.2 & \phn3.2 $\pm$ 0.2 \\
CH$_3$OH 44 GHz\tablenotemark{b}
               & 10.32 $\pm$ 0.08 & 1.3 & 0.8 $\pm$ 0.2 & \phn1.1 $\pm$ 0.3 \\
               & 11.32 $\pm$ 0.08 & 1.6 & 0.7 $\pm$ 0.2 & \phn1.2 $\pm$ 0.4 \\
               & 12.21 $\pm$ 0.14 & 0.9 & 0.7 $\pm$ 0.3 & \phn0.7 $\pm$ 0.3 \\
CH$_3$OH 95 GHz\tablenotemark{a}
               & 10.99 $\pm$ 0.13 & 2.5 & 1.8 $\pm$ 0.3 & \phn4.7 $\pm$ 0.7 \\
CH$_3$OH 133 GHz\tablenotemark{a}
               & 10.41 $\pm$ 0.19 & 7.7 & 1.7 $\pm$ 0.4 &    13.7 $\pm$ 2.7 \\
\enddata\\
\tablenotetext{a}{Single-component Gaussian fit.}%
\tablenotetext{b}{Triple-component Gaussian fit.}%
\end{deluxetable}

Considering the non-detection of the other CH$_3$OH lines
and the small width of each velocity component of the 44 GHz line,
we favor the interpretation that the detected lines are maser emission.
However, it is possible that the detected emission
can be a mixture of thermal and maser components
(e.g., Kalenskii et al. 2010).
Additional observations such as interferometric imaging
are needed to confirm the nature of this emission.

There is no map of the CH$_3$OH lines
with a sensitivity good enough to constrain the source position,
and the YSO(s) responsible for the maser emission
can only be deduced from the beam size.
Since these lines have different beam sizes (Table 2),
it is possible
that some emission structure in the beam of a low-frequency line
can be outside the beam of a high-frequency line.
It is difficult to specify the source of the 44 GHz maser line
because there are several embedded objects (FIR 5w/e, 6c/n, and 7)
and other YSOs within the beam
(Barnes et al. 1989; Skinner et al. 2003; Rodr{\'\i}guez et al. 2003;
also see Figure 1(b) of Choi et al. 2012).
For the 95 and 133 GHz maser lines,
the source is most likely associated with FIR 6c/n
because the angular separations of FIR 5 and 7 from FIR 6c
($\sim$25$''$ and 21$''$, respectively)
are larger than or comparable to the beam sizes.
Note that IRS 22 (Class I YSO; Comer{\'o}n et al. 1996)
is located within the beam.

\subsection{The $^{13}$CS $J$ = 2 $\rightarrow$ 1 Line}

The $^{13}$CS line was observed
to measure the systemic velocity of the dense cloud core containing FIR 6.
Figure 6 shows the spectrum.
The best-fit Gaussian profile gives
$V_0$ = 10.77 $\pm$ 0.05 km s$^{-1}$, peak $T_{\rm mb}$ = 1.95 $\pm$ 0.03 K,
$\Delta V$ = 1.66 $\pm$ 0.11 km s$^{-1}$,
and $\int T_{\rm mb} dV$ = 3.44 $\pm$ 0.19 K km s$^{-1}$.

\begin{figure}[!b]
\epsscale{1.0}
\plotone{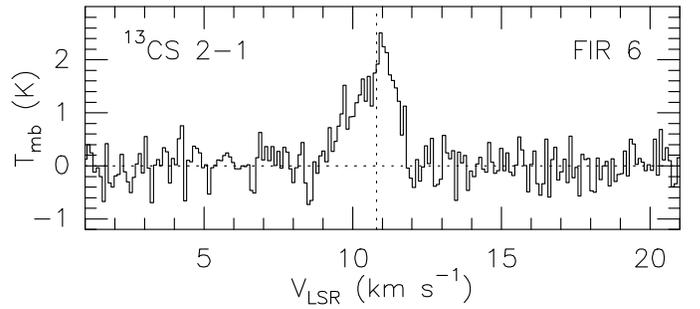}
\caption{\small\baselineskip=0.825\baselineskip
Spectrum of the $^{13}$CS $J$ = 2 $\rightarrow$ 1 line toward FIR 6.
Vertical dotted line:
centroid velocity of the best-fit Gaussian profile
($V_{\rm LSR}$ = 10.8 km s$^{-1}$).}
\end{figure}

The systemic velocity of the FIR 6 dense core has been known somewhat poorly
because the values in the literature were measured
using lines with relatively higher optical depths.
Schulz et al. (1991) derived 11.3 km s$^{-1}$
using the CS/C$^{34}$S $J$ = 5 $\rightarrow$ 4 and 7 $\rightarrow$ 6 lines
showing widths of 2.5--2.9 km s$^{-1}$.
Mangum et al. (1999) derived 10.6--10.9 km s$^{-1}$
using the H$_2$CO lines showing widths of 2.3--2.5 km s$^{-1}$.
For the derivation of systemic velocity,
$^{13}$CS is a better tracer than C$^{34}$S
because of its smaller abundance (Shirley et al. 2003).
Indeed, the $^{13}$CS $J$ = 2 $\rightarrow$ 1 line
is much narrower than the lines listed above.
Therefore, the systemic velocity of 10.8 km s$^{-1}$
derived from the $^{13}$CS line may be more reliable.

\section{DISCUSSION}

\subsection{FIR 6}

The detection of the H$_2$O and CH$_3$OH masers suggests
that the NGC 2024 FIR 6 region is actively forming stars,
which produces shocks in the warm and dense molecular gas
around the YSOs.
The two maser species show the well-known contrasting characteristics
(Menten 1991; Elitzur 1992).
The H$_2$O maser emission can be seen
in a relatively wide range of velocities
while the CH$_3$OH maser emission is confined
in a narrow range around the systemic velocity of the cloud.
The H$_2$O maser also displays
strong variabilities in both amplitude and velocity,
while the CH$_3$OH masers show no detectable variability.

The most likely exciting source of the H$_2$O maser reported in this paper
is FIR 6n.
The maser probably comes from the base of the bipolar outflow
driven by FIR 6n (Richer 1990; Choi et al. 2012).
During the observation period covered in this study,
all the H$_2$O emission was blueshifted
with respect to the systemic velocity of the FIR 6 core,
and the blueshifted outflow may have been
responsible for the observed H$_2$O maser emission.
However, the emission was nearly at the systemic velocity
in 1995/1996 (Choi et al. 2012),
and both blueshifted and redshifted velocity components
were detected in 1999 (Furuya et al. 2003).

The variability of the FIR 6 H$_2$O maser is typical of low-mass YSO.
The typical life time of each velocity component is $\sim$8 months.
This timescale is generally consistent with the values
reported in surveys toward low-mass YSOs,
which range from $\sim$2 weeks to $\sim$8 months
(Wilking et al. 1994; Furuya et al. 2003).
The velocity drift rate of the FIR 6 H$_2$O maser
is similar to that reported by Tofani et al. (1995).
Note that the time interval between observing runs in this study
is not short enough to trace variabilities
at timescales shorter than about a month.

The interpretation of the CH$_3$OH class I masers is unclear.
Soon after the discovery of the CH$_3$OH masers,
it was found that class I masers trace outflows driven by high-mass YSOs
and are located away from the responsible YSOs up to $\sim$2 pc
(Plambeck \& Menten 1990; Kurtz et al. 2004).
Recently, however, it was found
that some class I masers may be associated with shocks
driven into molecular clouds by expanding H {\small II} regions
(Voronkov et al. 2010; Chen et al. 2011)
and that several low-mass star forming regions emit class I masers
(Kalenskii et al. 2010).
If the FIR 6 CH$_3$OH masers trace an expanding H {\small II} region,
the FIR 6c hypercompact H {\small II} region (candidate) may be responsible.
If they trace an outflow,
the FIR 6n bipolar outflow may be responsible for the masers.
If the maser emission comes from the FIR 6n outflow,
it would be one of the rare examples of CH$_3$OH maser
associated with low-mass YSOs.
This issue can be settled by future observations using interferometers.

\subsection{FIR 4}

The detection of the H$_2$O maser toward FIR 4 illustrates
the importance of mapping in single-dish observations of H$_2$O masers.
While this paper gives the first confirmation
of the H$_2$O maser associated with FIR 4,
it is not necessarily the first detection.
For example, Furuya et al. (2003) reported
single-dish observations toward FIR 5/6 (see their Figure 13(a)).
Since FIR 4--6 are within their beam, it is likely
that the detected emission in their spectra were mostly from FIR 4/6.
(Note that interferometric observations
never showed an H$_2$O maser associated with FIR 5.)
Our recent survey also confirmed
that multiple H$_2$O maser sources in a single-dish beam
is a common occurrence in cluster-forming regions
such as the Orion cloud (M. Kang et al. 2012, in preparation).
The exciting source of H$_2$O maser emission in a given detection
is not necessarily the previously known source,
the youngest YSO in the beam,
or the most powerful outflow source in the beam.
This problem can affect some statistical quantities
derived from single-dish surveys, such as detection rates,
and can be worse in massive star forming regions far away from the Sun.

The H$_2$O maser suggests
that there are shocks produced by the star formation activities of FIR 4.
The nature of FIR 4, however, has been controversial.
Early studies indicated the existence of dense gas in the NGC 2024 region
(Evans et al. 1987).
Mezger et al. (1988, 1992) suggested
that the FIR 1--7 cores are isothermal protostars
without luminous stellar objects, based on their dust continuum observations.
Studies with molecular (thermal) lines suggested
that they contain more evolved YSOs (Schulz et al. 1991; Mangum et al. 1999).
The discovery of a near-IR reflection nebula and a CO outflow revealed
that FIR 4 contains a low-mass YSO driving a bipolar outflow
(Moore \& Chandler 1989; Moore \& Yamashita 1995; Chandler \& Carlstrom 1996).
A very different view was proposed by Minier et al. (2003).
They detected the CH$_3$OH class II maser line at 6.7 GHz in emission
and argued that FIR 4 contains an intermediate or high-mass YSO.
Future investigations are needed to tell
whether low-mass protostars can produce CH$_3$OH class II maser emission
or FIR 4 is a site of high-mass star formation.
The detection of the H$_2$O maser
does not necessarily favor one or the other possibility.

It is interesting to note
that there are position offsets of $\sim$2$''$
among the 3 mm continuum, CH$_3$OH class II maser, and hard X-ray positions
(Chandler \& Carlstrom 1996; Minier et al. 2003; Skinner et al. 2003).
Future determinations of the accurate position of the H$_2$O maser source
will be helpful in understanding the nature of FIR 4.

\section{SUMMARY}

The NGC 2024 FIR 6 region was observed
using the KVN in the H$_2$O 22 GHz maser line,
the CH$_3$OH 44, 95, and 133 GHz class I maser lines,
other CH$_3$OH lines in the tuning ranges,
and the $^{13}$CS $J$ = 2 $\rightarrow$ 1 line
in the single-dish telescope mode.
In addition, the FIR 4 region was observed in the H$_2$O line.
The main results are summarized as follows.

1.
Several velocity components were identified in the H$_2$O spectra.
The H$_2$O maps revealed
that most of them seem to be associated with FIR 6n
while one is associated with FIR 4.
None is associated with FIR 5.

2.
Month-scale variabilities of the H$_2$O maser were seen clearly.
A typical life time of each velocity component is $\sim$8 months.
The centroid velocity fluctuates
with a drift rate of $\sim$0.01 km s$^{-1}$ day$^{-1}$.

3.
The CH$_3$OH class I maser lines were detected toward FIR 6.
They were interpreted as maser emission.
The maser emission is confined
within $\sim$2 km s$^{-1}$ around the systemic velocity of the cloud.
It is not clear whether the exciting source is FIR 6c or 6n.
The CH$_3$OH masers did not show a detectable time-variability.
High-resolution imaging is needed to understand
whether the CH$_3$OH maser emitting region is shocked
by an outflow or an expanding H {\small II} region.

4.
The other CH$_3$OH lines, thermal or class II maser, were not detected.

5.
The $^{13}$CS spectrum showed
that the systemic velocity of the FIR 6 dense core is $\sim$10.8 km s$^{-1}$.

\acknowledgements

We thank the KVN staffs for their support.
M.C. and M.K. were supported by the Core Research Program
of the National Research Foundation of Korea (NRF)
funded by the Ministry of Education, Science and Technology (MEST)
of the Korean government (grant number 2012-0005532).
J.-E.L. was supported by the Basic Science Research Program
through NRF funded by MEST (grant number 2012-0002330).

\end{document}